# Design and Assessment for Hybrid Courses: Insights and Overviews

Felix G. Hamza-Lup
Computer Science and Information Technology
Armstrong State University
Savannah, Georgia, USA
Felix.Hamza-Lup@armstrong.edu

Stephen White
School of Human & Health Sciences
University of Huddersfield
Huddersfield, UK
Stephen.White@hud.ac.uk

*Abstract* - Technology is influencing education, providing new delivery and assessment models. A combination between online and traditional course, the hybrid (blended) course, may present a solution with many benefits as it provides a gradual transition towards technology enabled education. This research work provides a set of definitions for several course delivery approaches, and evaluates five years of data from a course that has been converted from traditional face-to-face delivery, to hybrid delivery. The collected experimental data proves that the revised course, in the hybrid delivery mode, is at least as good, if not better, than it previously was and it provides some benefits in terms of student retention.

I. INTRODUCTION

The abundance of computing power and the widespread availability of the Internet had a tremendous impact on society for the past decades. Education as a fundamental branch of social activity has been rapidly reshaping itself adapting to the informational era. The online teaching technology, like any novel approach, presents both advantages and disadvantages, appropriate use and misuse. This article presents the recent evolution of teaching styles vis-a-vis technology focusing on defining and assessing hybrid delivery methods. The research builds on previous work [1], adds additional related work studies, as well as an additional year of experimental data and discussions.

Studies in the United States (i.e., The Gartner Group Research Institute) anticipated that the world's e-Learning sales would grow 14.5% annually from 2006 to 2011 [2]. Over a similar timescale, government policies in the United Kingdom also indicated that the effective use of technology-assisted student-focused learning is essential for the future of higher education [3-6]. In a review of higher education and the future role of the university, Ernst & Young [7] have suggested that "… campuses will remain, but digital technologies will transform the way education is delivered and accessed, and the way 'value' is created by higher education providers, public and private alike."

Large scale as well as smarter use of technology in teaching is widely seen as a promising way of controlling costs [8]. When compared to other service industries, higher education stands out as being particularly affected by what has been described as the "cost disease" [9]. Universities have large costs for infrastructure and labor, with reliance on expensive face-to-face provision. The urgent need to boost university productivity has been noted by many [10-12]. Moreover, cost reductions are demanded by students (as they want to spend less time and lower the costs of traveling to the main campus) and improved time flexibility, specifically for full-time or part-time working students.

Face-to-face lectures are accepted as being a very inexpensive way of presenting new ideas and concepts to students. Additionally, lecturing has been described as an ineffective tool for promoting theoretical understanding [13], as it rarely stimulates student thinking beyond the short-term memory [14][15]. The passive role assumed by students in lectures is too focused on the subject being delivered, rather than the learners and their individual needs [16]. But, teaching the same content can be made more interesting, and students can become active, independent learners, if different delivery methods (including multimedia) are used [17].

Implemented proficiently, the online and the hybrid (blended) provision has the capacity to lower costs and at least sustain, if not boost student outcomes [18-20]. Hybrid/blended learning can ease some of the economic strain on students, as it reduces commuting expenses and allows for a flexible timetable that may better accommodate the students' personal circumstances [21]. Cost simulations, although speculative, have indicated that adopting hybrid models of instruction in large introductory courses has the potential to reduce costs quite substantially [8].

This article presents in Section II a set of definitions for the terms in common use in educational delivery, and provides clarifications on the use and meaning of these terms. The choice for hybrid/blended learning is described in Section III, followed by the "Fundamentals of the Internet and the World Wide Web" (CSCI 1150) course description in Section IV. The methodology for data collection is detailed in Section V, with Section VI exploring the evaluation of said data in terms of student outcomes and attrition rates. The relationship between assessment weighting and online student interactions in discussion forums is also measured and analyzed. Section VII identifies the limitations of this study and concludes that the CSCI 1150 course, in hybrid delivery mode, continues to provide as good, if not a better provision, than the previous traditional face-to-face delivery method.

## II. UNDERSTANDING HYBRID (BLENDED) LEARNING

The growth of e-Learning has blurred the boundaries of educational modes [22]. Higher education institutions and academics use a wide range of terms to describe ways in which students may engage with their studies, including on-campus, face-to-face, off-campus, open education, distance education, external study, online education, e-Learning, flexible learning, blended learning and hybrid. Both Lund and Volet [23] and Schlosser and Simonson [24] have suggested that there is limited consensus on the meanings of these terms, and a degree of confusion for academics, administrators and students exists within the university sector. For each learning environment listed above, there are distinct attributes that help locate and define them in a typological structure. For example, an on-campus mode relates to "courses that deliver material face-to-face and students interact with instructors face-to-face" [25], whilst distance learning (or education) is the various forms of study at all levels which are not under the continuous, immediate supervision of instructors collocated with their students. These forms of study, nevertheless, benefit from the planning and guidance of a supporting educational organization (Holmberg, cited in [24]). Still considering the location of delivery, Howland and Moore [26] suggest that an online course is "one in which no more than one face-to-face meeting is required". Offering a slightly contrasting view, Bollinger and Wasilik [27] consider a course to be online "if 80% or more of the content is delivered via the Internet". With the advent of technology in education, the boundaries of what learning environment fits within a mode of enrolment fades and misunderstandings arise.

While initial observations of computer-based learning have noted that "e-Learning is a confused and confusing field, fragmented into multiple disciplines and emphases" [23], a general definition is provided by Pollard and Hillage [28] who suggest e-Learning represents "the delivery and administration of learning opportunities and support via computer, networked and web-based technology to help individual performance and development". Kruse defines it as the use of technology to deliver learning programs and training programs through CD-ROMs, the Internet, local area networks (LANs), and wireless (WiFi) networks to promote active learning [29]. E-Learning and Computer-based Learning can be seen as broader than Online Learning as they do not always require web-based connectivity [30] since learning activities can occur on stand-alone devices.

The term Blended Learning is being used with increasing frequency in academic writing but there is no consensus on its meaning [31]. An alternative term, Hybrid, is defined as being of "mixed character; composed of different elements" [32], and Blended is defined as "an unobtrusive or harmonious part of a greater whole" [33]. Blended Learning has been described as a hybrid instructional approach combining aspects of e-Learning and a traditional classroom environment [30] and defined as "courses that deliver material both face-to-face and online" and where "students interact with instructors both online and face-to-face" [25]. Many colleges offer hybrid courses, which combine traditional face-to-face with online instruction. Previous research proves that this combination may promote learner-centered and active learning [34].

To understand the position and better define Hybrid (or Blended) learning we are looking at the possible categorizations of instructor-student interactions on different dimensions, e.g., space and time. Hence, instructor-student interaction can be categorized based on the geographical location/space as:

- Local (face-to-face): the instructor and the students share the same physical location, usually the classroom on the university campus grounds.

- Remote (distance): the instructor and the students do not share the same physical location. Students in this case can conveniently attend courses from home - the "living room" vs. the "classroom" paradigm.

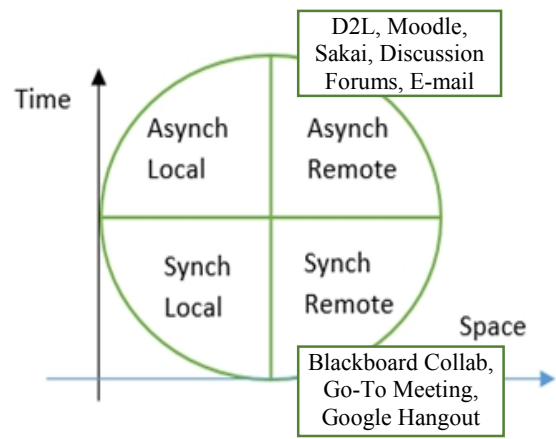

Figure 1. Time-distance diagram: local vs remote, synchronous vs asynchronous and applications used for the remote case.

The second dimension and an important categorization used for learning methods is based on the interaction time (or interaction style) among the course participants: students and instructor, as illustrated in Figure 1.

For synchronous interaction, the students and instructor are present and interact at the same time. Such interaction can occur in the traditional face-to-face setup or in a remote setup. Anohina [35] describes synchronous online learning as a method to bring a learning community together at the same time without distance being a barrier to interactions. Time flexibility can be an issue here. Web conferencing applications such as Blackboard Collaborate, Citrix Go-To Meeting, Adobe Connect, Google Hangout, and Videoconferencing are used in such synchronous interactions.

In case of asynchronous interaction, the students and instructor are not present at the same time. This type of interaction is not common in a traditional face-to-face setup but it is more common in a remote (distance) setup. It brings together the learning community without distance or time being a barrier to interaction. Learning management systems such as Desire-to-Learn (D2L), Blackboard, Moodle, Sakai, and others are used to deliver asynchronous online learning using a variety of tools (e.g., discussion forums).

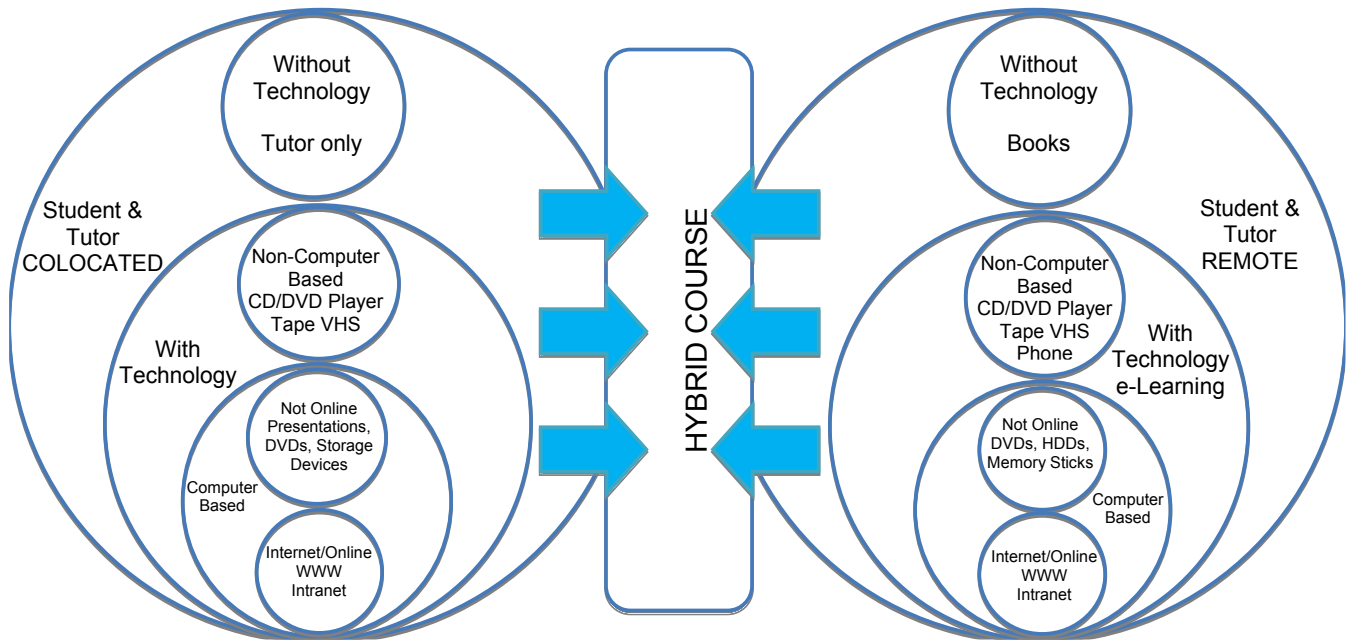

Figure 2. Hybrid courses in relation to traditional and online delivery

A third dimension is possible based on the use of a computing system: computer-based learning (CBL) uses software running from DVDs, remote servers (the "cloud") or downloaded to student computer for instruction. Many textbooks now come with CBL modules, which can stand alone or be incorporated into online course delivery. Non-computer based learning involves any device that is not a computer (e.g., DVD/video player, other devices).

From the above categorizations, a hybrid course (as illustrated in Figure 2) is defined as a course in which the instructor and student is partially remote and partially in the same physical location. A hybrid course does not necessarily use a computer and the Internet but, with the wide spread of these technologies, we find that it is very common for the hybrid course to be computer and Internet based.

Research shows that the hybrid mixture of on-campus and off-campus activities is difficult to explain to prospective students [36]. A potential solution to the confusion is to define courses specifically by their construction. The public University System of Georgia (USG) [37] defines the following course types:

- Fully online: All or nearly all the class sessions are delivered via technology (96% to 100% online).
- Partially online: Technology is used to deliver more than 50% of class sessions (51% to 95% online).
- Hybrid: Technology is used to deliver at least one class session and up to 50% of class sessions.
- Campus (or on-site): No class sessions are replaced by online technology.

The relationship between traditional, online, and hybrid courses, is displayed in Figure 2. Armstrong State University part of USG, defines three types of programs based on the level of online interaction [38]:

- Online program: includes only fully online courses. Does not included partially online or hybrid courses. Fully online programs are meant for those who live far from campus or may have jobs that prevent them from attending campus classes.
- Blended program: includes partially online and fully online courses. Blended programs are ideal for students who live on or near campus but work part-time.
- Campus/On-site program: this program is ideal for students living on or near campus who attend class part-time or full-time. These programs can be ideal for full-time students who benefit from the structure of face-to-face instruction. There are three types of campus/on-site courses: hybrid courses, technology-enhanced courses and "no technology" courses.

### III. THE CHOICE FOR HYBRID LEARNING

During lectures students usually assume passive roles as listeners while the instructor distributes information. Educating in this way is too focused on what is being delivered, rather than the learners and their needs.

White et al. [39] demonstrated that traditionally delivered, subject-intense courses can be converted to a 'blended/hybrid' delivery approach with "as good, if not better, outcomes", if they are well-designed with high quality content and regular interaction.

As observed in the e-Learning Ladder [16], a constructivist theory, the student, rather than the instructor, should be the focus, and he must 'construct' new knowledge through analysis, experience and understanding. The Ladder further identifies that opportunities for learners to be active in creating their own knowledge and understanding can be offered through Web 2.0, and more recently Web 3.0 technologies, such as: discussion boards or forums and various types of social media applications. These applications allow students to retrieve information, as well as provide a platform to create and own the data within them. These tools can be used as an alternative or in addition to traditional lectures resulting in a learner-centered environment [40].

There are also indirect benefits in using technology-enhanced learning, such as the development of students' computer skills. However, this is directly relevant to one problem commonly associated with e-Learning, because just as with any genre of course, learners need to have the appropriate resources in order to be successful. These resources in an e-Learning context can be classified as 'External' to the learner, such as slow Internet connections or older computers, and 'Internal' to the learner, which may be a lack of the necessary computer skills. Without these resources, accessing the course materials can be difficult and the learners' performance can be significantly hindered [41]. Anxiety can set in leading to lack of motivation [42], which may ultimately result in students becoming frustrated and giving up on that particular learning environment [43]. However, it has also been identified that an initial lack of confidence can be quickly replaced by positive excitement once the initial experience of e-Learning has taken place and the technology involved mastered [44].

The goal of a blended learning experience is "to provide a mix of both on-line and face-to-face experiences which support each other in achieving desired learning outcomes" [45]. Many universities experiment with a blended learning model as part of their teaching strategy, but "the term is still relatively new, therefore, leaving many to question how the mixing of online and mobile learning with face-to-face interaction will actually improve student experience now and in the long term" [46]. This hybrid mixture of on-campus and off-campus activities [36] is difficult to explain to prospective students.

Combined with the need to be self-motivated and more independent, hybrid learning is most definitely no easier than the face-to-face course. Nor is it for all students. For this reason, among others, it is imperative that hybrid learning be carefully crafted from a pedagogical standpoint. That begins with the effective definition of the course goals and objectives. Goals are broad, generalized statements about what is to be learned, and they can be taught of as targets to be reached. The objectives are the base upon which one can build lessons and assessments that meet the overall course goals. From this solid foundation, the course content and student assessment has to be designed and implemented in a completely different fashion than for the traditional face-to-face course.

Bowen et al. [8] aimed to estimate the costs associated with course delivery under different circumstances. Whilst acknowledging that the simulations are admittedly speculative in nature and subject to considerable variation depending on how a particular campus organizes its teaching, they suggest that significant cost savings are possible. In particular, they estimate savings in compensation costs for the hybrid model ranging from 36% to 57% compared to the traditional model. These simulations confirm that hybrid learning offers opportunities for significant savings, but the degree of cost reduction depends on exactly how hybrid learning is implemented, especially the rate at which instructors are compensated and section size. A large share of cost savings is derived from shifting away from time spent by expensive professors toward computer-guided instruction. Their simulations substantially underestimated the savings from moving toward a hybrid model in many settings, because they did not account for space costs. It is difficult to put a dollar figure on space costs because capital costs are difficult to apportion accurately to specific courses, but the difference in face-to-face meeting time implies that the hybrid course requires 67% to 75% less classroom use than the traditional course. In the short run, institutions cannot lay off tenured faculty or sell or demolish their buildings. In the long run, however, using hybrid models for some large introductory courses would allow institutions to expand enrolment without a commensurate increase in space costs, a major savings relative to what institutions would have to spend to serve the same number of students with a traditional model of instruction. In other words, the hybrid model need not just "save money"; it can also support an increase in access to higher education. It serves the access goal, both by making it more affordable for the institution to enroll more students, and by accommodating more students because of greater time and space scheduling flexibility.

IV. HYBRID COURSE DESIGN

The course that we evaluate, CSCI 1150, had traditionally been taught face-to-face, in both the spring and fall semesters of 2010. In 2011, a Desire-to-Learn (D2L) online version of the course was developed. D2L is a web-based course management system that students were already familiar with. The content was made available online, within PDF slides that closely followed the associated textbook. The main reason of the slides approach was to provide structural guidance for the content in the textbook. Students were also provided with access to various interaction tools, both synchronous and asynchronous (e.g., e-mail, chat, discussion forums) as well as a set of assessment tools (e.g., online quizzes, online assignments and online exams).

The course content has been refined in subsequent years from 2012 to 2015, to include additional required reading material, as well as a better-defined set of discussion forums (one discussion forum per textbook chapter), where students were encouraged to interact during the semester. This refinement aimed to provide fresh stimuli to the course, in order to promote students' learning through questioning, investigating, challenging, seeking feedback, and learning through interactions with peers and tutors [47]. Technologies such as discussion forums can provide the opportunity for learners to be active in creating their own knowledge and understanding by allowing them to create, own, retrieve and exchange information within them [48] in a time flexible manner. The face-to-face sessions were then used to explore the course content, and the online interactions, in order to further develop the students' understanding. This overall course design may be seen as consistent with the "flipped classroom" approach [49], and is presented in a 50:50 ratio, causing it to be described as Hybrid delivery under the University System of Georgia definition [37].

From spring 2012, the course assessment has also been completed online, with each element assigned a proportion of the overall grade: Assignments were weighted at 40%, Quizzes weighted at 10%, the Midterm exam at 25% and the Final exam at 25%. The grading scheme was further supplemented from fall 2013, with the online forum interactions being rewarded 2% of the weight, reducing the Midterm and Final exams to a weight of 24% each. The online interaction (based on discussion forums) weight has subsequently been increased to 10% of the final grade for the spring semester of 2014, causing the Midterm and Final exams weights to be reduced to 20% each. For the spring 2015 semester, the online interaction weight was further increased to 15% of the final grade by reducing the quizzes weight from 10% to 5%. These adjustments were driven by the need to increase student-to-student interaction, and to measure how increasing reward affects that interaction as well as student learning in a hybrid setup.

*A. Automatic vs. Manual Grading*

A course management system like D2L provides advantages to both the faculty and the student. It is possible to automate the process of quiz and exam delivery as well as grading, subsequently freeing significant faculty time and providing timely feedback to students. Freeing faculty time is of upmost importance in a current complex higher education system that requires teaching, service as well as research activities from the faculty.

The online quizzes for the hybrid course were designed to be administered quickly (on average they are timed to 30 minutes each) and contained about ten questions each. The questions were automatically and randomly selected from a database of 3000+ questions, all of the same difficulty level. The quizzes were automatically graded, immediately after the deadline, providing students with instant access to both the grade and the correct solutions. Then, students can use this information to identify where they went wrong and what concepts they misunderstood in each chapter. The results of the quizzes was then discussed both online and in the face-to-face session, both with the instructor and student peers.

The drawback in automating the process of delivery and grading comes from the fact that some type of problems, such as those requiring essay-type answers, are difficult to automate, as they require manual grading for optimum accuracy, as well as to provide constructive personalized feedback. For this reason, the manually graded assessment components have a greater weighting in the overall final grade and includes personalized detailed feedback from the instructor for each student.

*B. Deadlines and Penalties*

Each assessment component has strict completion deadlines. Assignments have to be completed in three weeks, with a deadline enforced through the D2L submission system. Late submission was not accepted, and failure to submit an assignment would almost certainly result in dropping a grade point, as the assignment weight was 10% of the final grade.

For the quizzes, each weighted at only 1% (respectively 0.5% for the spring of 2015) of the final grade, there is a two to three weeks' timeframe during which each quiz can be taken, providing the students with time flexibility in their learning schedule, but keeping them on track with the rest of the course pace. Therefore, quizzes are designed to keep the students on track with their learning of the course content, providing them with early and progressive feedback concerning their course progress.

As previously identified, the Midterm and Final exams were also given online, with a strict 12-hours window where they are 'live' and can be taken. Each exam consists of 10 problems, with 80% of the responses being essay type, therefore, requiring manual grading. Each exam is weighted at 20%, with no late submissions allowed. Specific D2L technologies like the Respondus LockDown Browser™ mode as well as the tight time-frame is used to hinder student cheating. A LockDown Browser prevents users from being able to cheat by printing, copying or browsing other websites while taking the tests. Once the test has started, the computer remains locked until the test is completed and submitted for grading, hindering cheating attempts.

The final element of assessment, is based on the interactions among students in a set of discussion forums. Students are allowed to post information and ask questions for one month in each forum. After the expiration date (which is set and announced for each forum) the students can still read the posts, however, they cannot post new content. The ability to read the forums content is important since it provides students with a continuous source of information. Particularly for slower (or problem) students it is important to provide means to catch up with the rest of the class without disturbing the pace of the course for the entire class.

The discussion forums contributions were weighted at 10% of the final grade, with the posts content evaluated subjectively by the instructor; being measured both quantitatively and qualitatively. As mentioned for spring 2015, the online interaction weight through the forums was further increased to 15% of the final grade and the outcomes measured. Further increase in the discussions forums weight may be beneficial to student-student interaction and may positively affect student learning as some students are sometimes eager to explain the concepts they understood to their peers.

*C. Interaction*

Besides the face-to-face interactions in class, two types of written discussions are frequently used in a hybrid course: synchronous and asynchronous. Whereas synchronous discussion requires participants to log in at a predetermined time and simultaneously join the discussion, asynchronous activities allows users to organize, read, and post messages at their own pace, as dictated by their preferred schedule.

Where online/hybrid course designers have opted for the use of discussion forums, they play an important role, often making up the major part of the students' activities and providing evidence of attendance, class participation, and sometimes assessment [50-53]. The delayed element to asynchronous communication, can allow participants more time to consider their responses, promoting deeper consideration and reflection of the subject [54][55]. In spite of this, it has also been argued that scholarly thinking regarding assessment of online discussion has not kept pace with the growing popularity of such practices [56].

The asynchronous interactions in the CSCI 1150 hybrid course employ the e-Mail system, a News system, and the Discussion Forums, the latter consisting of one primary thread per textbook chapter. The News system is an efficient tool for the instructor to provide students with updates about the course, however, it is a unidirectional communication tool - from instructor to students.

Online synchronous interaction was implemented in CSCI 1150 through a Chat channel. It has been observed that the chat channel is mainly used immediately prior to the Midterm and Final exam period, serving as an emergency notification tool for the students if or when something goes wrong with the online exam session.

The other synchronous interaction occurred in the traditional in-class face-to-face meetings. As part of the Hybrid course, students meet with their instructor once a week, for a 75-minute session, where they can discuss and ask/answer questions. Attendance is not mandatory and it has been observed that by the middle of the semester only an average of 60% of the students attend these sessions mainly due to their part-time, full-time work schedule or other family commitments.

Online interaction through the spring and fall of 2014 was stimulated through the relationship between this activity and the assessment. Ten percent (10%) of the final grade was awarded for the discussion forum posts, with each student being expected to provide at least three posts per discussion thread, each being a paragraph of 200 words or more, as well as responding to classmates' questions providing original answers and/or alternate solutions. At the end of the semester, the student with the highest number of quality posts receives a further 10% towards his/her final grade; the other students receive lower additional percentages, representative of their contributions. A further increase to 15% of the final grade was implemented for spring 2015, as previously mentioned, and the effects of this assessment policy were measured and analyzed.

## V. METHODOLOGY

The CSCI 1150 course, "Fundamentals of the Internet and the World Wide Web", is a service course at Armstrong State University, Georgia, US. The course was observed over a period of four and a half years, through seven semesters (spring and fall, 2011 to spring 2015). Each semester consisted of two or three sections of the course hence 50 to 75 students were observed each semester (as further illustrated in Section VI). The course was delivered by traditional face-to-face methods in 2011, and was then converted to the Hybrid delivery mode for the 2012-2015 time frame. There is no entry, prerequisite requirement for the course.

The average class size was 25, and the students included in the data collection ranged from 19 to 42 years of age, with a female to male ratio of 1.7 to 1. The analysis of the experimental data is straightforward. The outcomes for students previously undertaking the course in the traditional face-to-face format are compared to the outcomes for students undertaking the hybrid format.

The data collected consists of the students' final grades, failure rates and withdrawal rates. To further evaluate the hybrid delivery method, the students' asynchronous interactions are also investigated. The rate and volume of posts in the online discussion forums are analyzed in consideration of the changes in the course structure and assessment strategy. The number of read post in the forums as well as the number of written contributions in the discussions forums are collected and analyzed in the light of the various grading weights imposed.

## VI. COURSE EVALUATION

The final outcomes for the students assessment are displayed in Figure 3, and these show no significant difference between the traditional face-to-face course that was delivered in 2011, and the subsequent hybrid delivery mode, with the course mean grade fluctuating between a B grade and a C grade (except for the anomalous D mean for the spring semester of 2011, course section 1).

TABLE I. MEAN AND MEDIAN GRADES FOR SECTIONS

|  | Mean Grade | Median Grade |
|---|---|---|
| Spring 2011 Section 1 | D | D |
| Spring 2011 Section 2 | C | C |
| Fall 2011 Section 1 | C | C |
| Fall 2011 Section 2 | B | B |
|  |  |  |
| Spring 2012 Section 1 | B | B |
| Spring 2012 Section 2 | C | C |
| Spring 2012 Section 3 | C | C |
| Fall 2012 Section 1 | C | B |
| Fall 2012 Section 2 | C | B |
| Spring 2013 Section 1 | C | C |
| Spring 2013 Section 2 | C | B |
| Spring 2013 Section 3 | B | B |
| Fall 2013 Section 1 | B | B |
| Fall 2013 Section 2 | C | C |
| Spring 2014 Section 1 | C | B |
| Spring 2014 Section 2 | B | B |
| Fall 2014 Section 1 | C | B |
| Fall 2014 Section 2 | C | B |
| Spring 2015 Section 1 | C | B |
| Spring 2015 Section 2 | C | B |

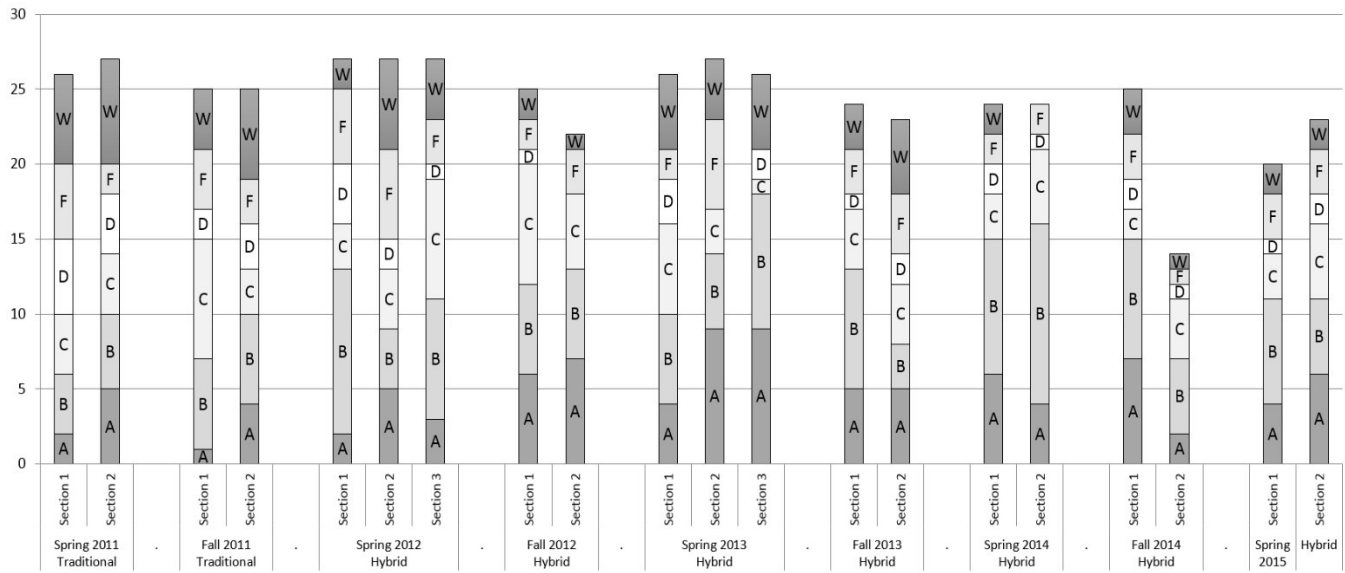

Figure 3. Total Number of Students, Number of Each Final Grade and Number of Withdrawals, per Section. (Grades A-D, F=Fail, W=Withdrawn)

There is, though, some suggestion, that the course outcomes may be improving, with a median of grade B appearing more regularly in the recent hybrid courses as illustrated in Table I; but whether this is due to the delivery method, or some external factor, cannot be determined precisely.

As mentioned earlier, the goal of a blended/hybrid learning experience is "to provide a mix of both on-line and face-to-face experiences which balance and support each other in achieving desired learning outcomes" [45]. Our results show that students taking the hybrid course format pay no "price" for this mode of instruction in terms of exam scores, and overall performance, proving that it is possible to hybridize certain courses without negative impact on learning. On the contrary, positive outcomes are possible for both the student and the instructor, resulting in time flexibility and cost savings on both sides.

In other sectors of the economy, the use of technology has increased productivity, measured as outputs divided by inputs, and has even often increased output. Bowen et al. [8] showed that a hybrid-learning system did not increase outputs (student learning) but could potentially increase productivity by using fewer inputs. When considering the course attrition rates, it is important to note that students are allowed to withdraw without penalty before an identified deadline – usually just after the Midterm exam. This allows failing students to leave with a "clean record", meaning they can retake the course in the future, should they wish to. Despite this, there is positive indication that attrition rates are reducing, as illustrated in Figure 4. However, this is unlikely to improve significantly under the current withdrawal policy.

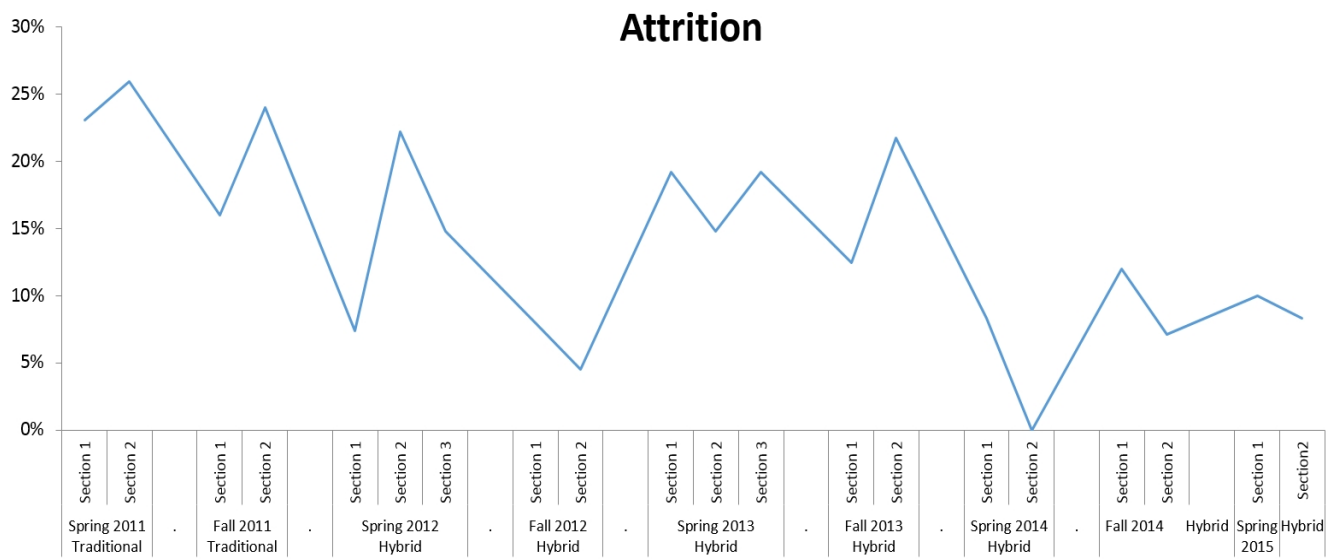

Figure 4. Course Attrition by Percentage of Total Enrolled Students (showing declining attrition)

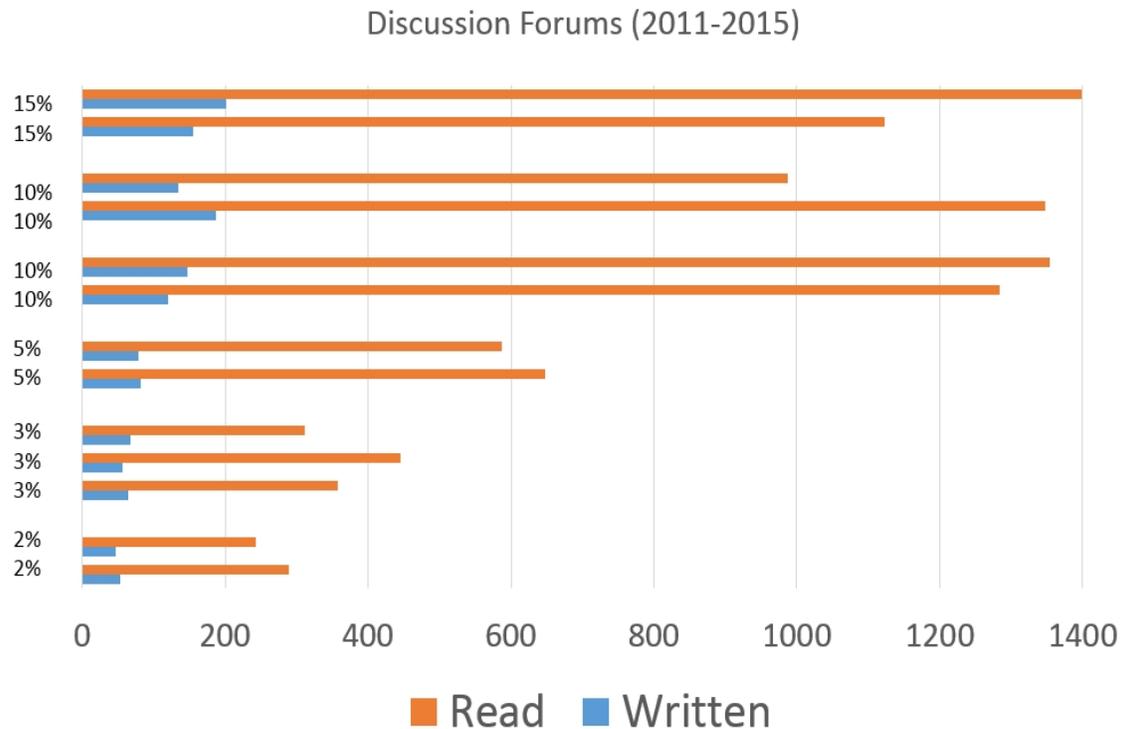

Figure 5. Relationship between the Forums Weight (2% to 15% of the Final Grade) and the Number of Read and Written Posts for each Semester, for each Course Section.

As previously identified, asynchronous interactions through the e-mail system are primarily exchanged around (one or two days, before and after) a major deadline for an assignment or exam. For example 76.5% of the e-mails received for sections 1 and 2 during the spring semester of 2014, were specifically targeted on questions around major assessment components. Students also tend to interact little amongst themselves using the e-mail system, with only 36% of the e-mails on average being sent for student-to-student communications for the same semester. For the spring semester of 2015 similar data was obtained: 74.4% of the e-mails received were specifically targeted on questions around assignments and exams. Student to student communication using the e-mail system amounted to approximately 34% of the total e-mails exchanged, showing that students prefer to communicate through other means (e.g., face-to-face, discussion forums etc.).

For the online interactions measured only through the discussion forums (from fall 2012 to spring 2015), a quantitative analysis of the forum contributions in terms of the number of written (authored) posts as well as the number of read posts reveals, somewhat unsurprisingly, that there is a direct dependency between the grading weight of the online interaction and the number of posts in the forum. Evidence shows that the higher the forum contribution weight is in the final grade, the higher the volume (and quality, in the instructor's opinion) of forum posts made by the students, as clearly illustrated in Figure 5. Specifically, one can observe a significant jump in the interest of "reading" the colleagues contributions, as well as a more moderate jump in the number of written contributions. This behavior could be explained by the fact that once the students use the forums, pushed by the grade constraint, they actually only then discover its contents value in solving assignment problems. This factor stimulates them to continue reading their colleagues posts and collaborate for problem solving. Stimulating such behavior, where student-student interaction is promoted, generates a superior learning environment where students become active learners. Active learning engages students in doing things and thinking about the things they are doing, becoming in this way problem solvers. A subjective, qualitative analysis of the forum posts advocates that students engage through these forums in higher-order thinking tasks such as analysis, synthesis, and evaluation.

VII. CONCLUSION AND FUTURE WORK

In this paper, Hybrid/Blended learning is discussed in the context of the existing terminology. The design, as well as the main components of a course that was transformed from a traditional face-to-face format to a hybrid one, is described.

The course analysis and evaluation focuses on the outcomes for students that undertook the course in the traditional format, and the outcomes for students undertaking the revised hybrid formats. We show that students in the hybrid format pay no "price" for this mode of instruction in terms of pass rates, exam scores, or performance. Moreover, they can be motivated to interact online with slight adjustments in the grading policy, which promotes participation, interaction and improves students' computer skills while, at the same time, engages them in active learning.

The evidence supports the hypothesis that well-designed interactive hybrid systems in higher education, have the potential to achieve at least equivalent if not better educational outcomes as traditional courses, while opening up the possibility of freeing up significant resources on both sides: student and instructor. These resources (e.g., time, classroom space and financial) could be redeployed more productively. This alone is cause for the hybrid style of course delivery to be recommended.

In spite of all these benefits caution must be taken when choosing courses to be hybridized. Not all the courses are fit for hybridization, moreover significant man-hours and resources may be required to develop the proper hybrid version for a course. Such cost considerations have not been investigated in this research and depend not only on the course content but also on the institution policy for course development.

An assumption, from the administrative point of view, is that the hybrid course sections can be larger and accommodate more students. In general this is a wrong assumption and also a dangerous one, since it creates the false illusion of immediate Full Time-Equivalent (FTE) improvement for faculty. Special care must be taken, as some studies [57] suggest that having 15 students per section is a good starting number for courses that have online exposure while their face-to-face equivalent would accommodate twice as much, around 25 to 30 students. Therefore, one should not rush into conclusions that hybrid mode delivery would automatically accommodate more students per class section.

As future work, the course structure will continue to be reviewed, in consideration of student outcomes, to promote higher final outcomes.

We would like to mention that this is a relatively limited-scale study and the data was drawn from a specific course, with a medium number of participants oscillating from 50 to 75 participants per semester, depending on the number of sections taught. The study may have been influenced by factors specific to the student groups, which are not immediately evident from the findings. Also, experiences external to the course content and delivery may have contributed to student outcomes and opinions. In the near future we will provide additional data as this hybrid course continues to be taught at our university.